\documentclass[aps,prb,twocolumn,superscriptaddress]{revtex4-2}

\usepackage{graphicx}
\usepackage{dcolumn}
\usepackage{bm}  
\usepackage{amssymb}  
\usepackage{amsmath}
\usepackage{epstopdf}
\usepackage[T1]{fontenc}
\usepackage{xcolor}
\usepackage{lmodern}
\usepackage[utf8]{inputenc}
\usepackage[english]{babel}
\usepackage{hyperref}
\hypersetup{colorlinks=true,citecolor={blue},linkcolor={blue},urlcolor={blue}}

\makeatletter 
\renewcommand{\fnum@figure}{FIG.~\thefigure}
\makeatother

\begin{document}

\title{Occupancy-driven Zeeman suppression and inversion in trapped polariton condensates}

\author{Krzysztof~Sawicki}
\email[]{k.sawicki@soton.ac.uk}
\affiliation{School of Physics and Astronomy, University of Southampton, Southampton SO17 1BJ, United Kingdom}
\author{Dmitriy~Dovzhenko} \affiliation{School of Physics and Astronomy, University of Southampton, Southampton SO17 1BJ, United Kingdom}
\author{Yuan~Wang} \affiliation{School of Physics and Astronomy, University of Southampton, Southampton SO17 1BJ, United Kingdom}
\author{Tamsin~Cookson} \affiliation{School of Physics and Astronomy, University of Southampton, Southampton SO17 1BJ, United Kingdom}
\author{Helgi~Sigur{\dh}sson} 
\affiliation{Institute of Experimental Physics, Faculty of Physics, University of Warsaw, ul.~Pasteura 5, PL-02-093 Warsaw, Poland}
\affiliation{Science Institute, University of Iceland, Dunhagi 3, IS-107 Reykjavik, Iceland} 
\author{Pavlos~G.~Lagoudakis} \affiliation{School of Physics and Astronomy, University of Southampton, Southampton SO17 1BJ, United Kingdom}
\vskip 0.25cm

\begin{abstract}
We study the magneto-photoluminescence of an optically trapped exciton-polariton condensate in a planar semiconductor microcavity with multiple $\mathrm{In}_{0.08}\mathrm{Ga}_{0.92}\mathrm{As}$ quantum wells. Extremely high condensate coherence time and continuous control over the polariton confinement are amongst the advantages provided by optical trapping. This allows us to resolve magnetically induced $\sim \mu$eV fine-energy shifts in the condensate, and identify unusual dynamical regions in its parameter space. We observe polariton Zeeman splitting and, in small traps with tight confinement, demonstrate its full parametric screening when the condensate density exceeds a critical value, reminiscent of the spin-Meissner effect. For larger optical traps, we observe a complete inversion in the Zeeman splitting as a function of power, underlining the importance of condensate confinement and interactions with its background reservoir excitons.
\end{abstract}

\maketitle

\section{Introduction}

Reconfigurable and highly nonlinear cavity-polariton systems controlled by external fields could be key for the development of flexible spinoptronic semiconductor devices~\cite{Liew_PhysE2011} and optical topological insulators and lasers~\cite{Solnyshkov_OptMatExp2021}. In contrast to weakly interactive pure photonic systems, a medley of active materials embedded in planar microcavities~\cite{Sanvitto_NatMat2016} has given unprecedented insight into strong light-matter physics under variable electric, magnetic, and optical fields~\cite{Yang_AdvQuaTech2022}. This feature results from the hybrid nature of exciton-polaritons (hereinafter called polaritons), which are formed by the strong interaction of light and matter and inherit the properties of both their components - photons and excitons (i.e., bound electron-hole pairs)~\cite{Carusotto:RMP2013}. Possessing extremely light effective mass and strong interactions, polaritons can accumulate in a non-equilibrium analogue of a Bose-Einstein condensate~\cite{Kasprzak:Nature2006} forming a strongly nonlinear polariton laser that can be electrically driven~\cite{Schneider_Nature2013, Bhattacharya2014} at both cryogenic and room temperatures.

Spinor polaritons possess two integer spin projections $s = \pm 1$ on the growth axis of the cavity, explicitly related to the two circular polarizations $\sigma^\pm$ of emitted light~\cite{Carusotto:RMP2013}. Hence, when the polaritonic system is illuminated with a linearly polarized nonresonant pump, the condensate also becomes linearly polarized with a polarization vector determined by the cavity strain or anisotropic disorder. The situation changes when an external magnetic field is applied parallel to the growth axis (i.e., Faraday geometry). Due to the underlying electron and hole constituents in the exciton wavefunction, the magnetic field results in a Zeeman effect between the $|\psi_+\rangle$ and $|\psi_-\rangle$ polariton states denoted by the splitting,
$E_{\mathrm{ZS}} = E_+ - E_-$. The appearance of fine-structure energy splitting is a fundamental manifestation of the influence of the magnetic field on the polariton structure. In particular, the dynamical interplay between the real magnetic field and an effective magnetic field caused by spin-anisotropic polariton-polariton interactions can lead to the full 
parametric screening of the former~\cite{Rubo:PRA2006, Liew:PRB2008}. This effect, known as the polariton spin-Meissner effect, is a manifestation of collective quantum behaviour in driven-dissipative polariton fluids.

For more than a decade, the magnetic properties of polaritons in planar microcavities or micropillars attracted considerable attention, accumulating in the observation of the Zeeman and spin-Meissner effects~\cite{Liew:PRB2008, Larionov:PRL2010, Walker:PRL2011, Rahimi-Iman:PRB2011, Kulakovskii_PRB2012, Pietka:PRB2015, Sturm:PRB2015, Krol:SciRep2018, Krol:PRB2019, Whittaker:PRA2021, Mirek:PRB2023}. Moreover, investigations using elliptically polarized excitation led to an optical analogue of the Zeeman effect~\cite{Krizhanovskii:PRB2006, Askitopoulos_PRB2016, Yago_PRB2019, Gnusov_PRB2020, Real:PRR2021, Baryshev_PRL2022, Sigurdsson_PRL2022, Dovzhenko:PRB2023}. In recent years, the effect of a magnetic field on polaritons has also been studied in microcavities with semi-magnetic quantum wells reporting \textit{giant} Zeeman splitting~\cite{Krol:SciRep2018, Krol:PRB2019, Sciesiek:CommunMat2020, Mirek:PRB2023}. However, an inherent limitation of the systems studied so far has been their lack of reconfigurability, large disorder, and low coherence times, which restricts polaritons for potential spinoptronic applications~\cite{Liew_PhysE2011}.

Remarkable progress in the technology of the epitaxial growth of GaAs-based layers enables fabrication of high-quality optical microcavities exhibiting significantly low disorder, wherein optically trapped condensates~\cite{Askitopoulos:PRB2013, Cristofolini:PRL2013, Ohadi:PRX2015, Balas_PRL2022} display the self-induced Larmor precession with remarkably long spinor coherence times of up to $\sim$ 9 ns~\cite{Baryshev_PRL2022, Sigurdsson_PRL2022}, three orders of magnitude longer than the polariton lifetime. When dynamically driven, the spin coherence time can even reach hundreds of ns~\cite{Gnusov:arxiv2023}. However, so far, the influence of the magnetic field on optically trapped polariton condensates has not been investigated. In particular, spin-related phenomena such as the spin-Meissner effect, which are difficult to observe due to the relatively small Zeeman effect of GaAs and InGaAs quantum wells (QWs), have not been observed in optically trapped condensates.

In this article, we present experimental evidence of both the Zeeman and spin-Meissner effects in an optically trapped polariton condensate. Observation of the abovementioned effects is possible due to the decreased overlap between the pump induced background reservoir of incoherent excitons and the stimulated coherent polariton condensate, consequently decreasing the linewidth of the condensate emission. This makes it possible to detect even minor variations in the polariton fine energy structure. Moreover, using the all-optical reconfigurability of the investigated system, we show that we move from suppressed Zeeman splitting to inverted splitting by simply tuning the optical trap size and the condensate density. Our findings are supported by a generalized spinor Gross-Pitaevskii equation coupled to a reservoir rate equation.

\section{\label{exp_details} EXPERIMENTAL DETAILS}

The studied sample is a strain-compensated, high-$Q$ ($Q\sim12000$~\cite{Cilibrizzi:APL2014}) GaAs-based 2$\lambda$ microcavity with embedded $6\,\mathrm{nm}$ $\mathrm{In}_{0.08}\mathrm{Ga}_{0.92}\mathrm{As}$ QWs (see schematic Fig.~\ref{KSawicki_Fig1_scheme}). In the cavity region, three pairs of QWs are located in the central three anti-nodes of the electric field. Two additional QWs positioned at the extreme nodes of the cavity wells serve for carrier collection. The top (bottom) distributed Bragg reflector is made of 23 (26) pairs of alternating refractive index $\mathrm{GaAs}$ and $\mathrm{AlAs}_{0.98}\mathrm{P}_{0.02}$ layers. The microcavity is of a wedge type, which allows tuning of the light and matter fractions of polaritons by choosing the  appropriate in-plane location on the sample. In this study, the experimental measurements are conducted at the exciton-cavity mode detuning of around $-1.9\,\mathrm{meV}$ in the absence of the magnetic field.

The experiments are performed at cryogenic temperature ($\sim7\,\mathrm{K}$) in a closed-cycle cryostat equipped with a superconducting magnet producing a magnetic field parallel to the optical axis in the range from $-5$ to $5\,\mathrm{T}$. The sample is excited non-resonantly with a linearly polarized continuous-wave Ti:Sapphire laser tuned to the Bragg reflector's minimum ($\lambda_{\mathrm{exc}}=758.8\,\mathrm{nm}$) of the sample and modulated by an acousto-optic modulator to avoid heating. The pump beam excites co-localized high-energy charge carrier distribution, which undergoes fast energy relaxation to form an incoherent exciton reservoir which in-turn feeds the condensate~\cite{Carusotto:RMP2013}. Because the photoexcited reservoir of excitons produces not only local gain for polaritons but also local blueshift, the resultant condensate can be confined when the laser beam is structured into an annular profile~\cite{Askitopoulos:PRB2013}. For this purpose, we use a spatial light modulator to create a ring-shaped pumping beam, similar as in the case of a set of axicon lenses~\cite{Manek:OpticsCommunications1998}, which forms the transverse trap. The photoluminescence (PL) is collected through a microscope with a numerical aperture of $0.4$, allowing for the collection of light from the condensate within the trap and the surrounding exciton reservoir. The combination of a quarter-wave plate and a Wollaston prism is used to simultaneously detect signals from both circular polarization components.

\section{EXPERIMENTAL RESULTS}
\subsection{Zeeman splitting of optically trapped polaritons}

\begin{figure}[t!]
\includegraphics[width=0.8\columnwidth]{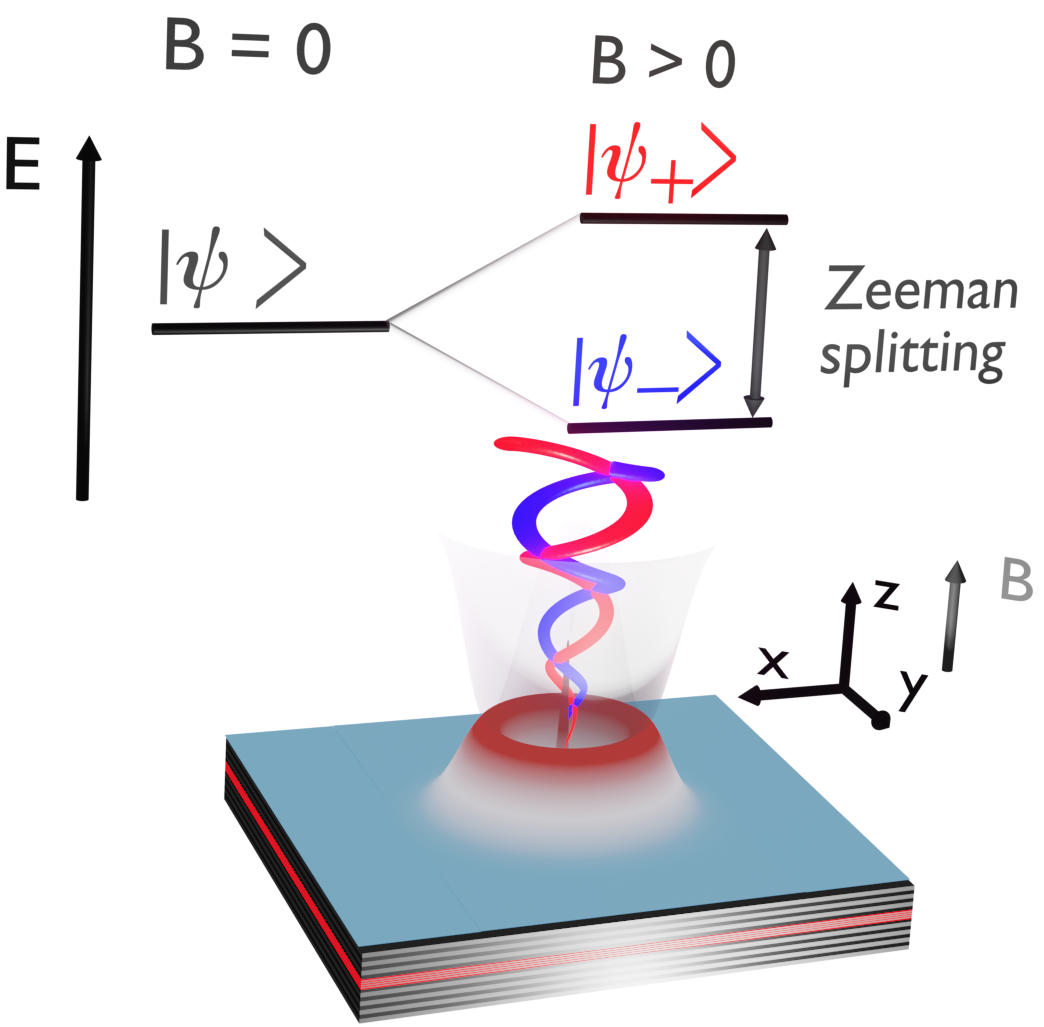}
\caption{\label{KSawicki_Fig1_scheme} Schematic illustration of the investigated sample. The linearly polarized, non-resonant continuous wave laser was used to create equal populations of the $|\psi_\pm \rangle$ polaritons. The corresponding $\sigma^\pm$ emission (blue and red spirals) is detected simultaneously. The magnetic field applied parallel to the sample growth axis lifts the degeneracy of the polariton spins, manifesting in a detectable energy difference (i.e., Zeeman splitting) between the emitted circularly polarized photons.}
\end{figure}

\begin{figure*}[t!]
\includegraphics[width=0.7\linewidth]{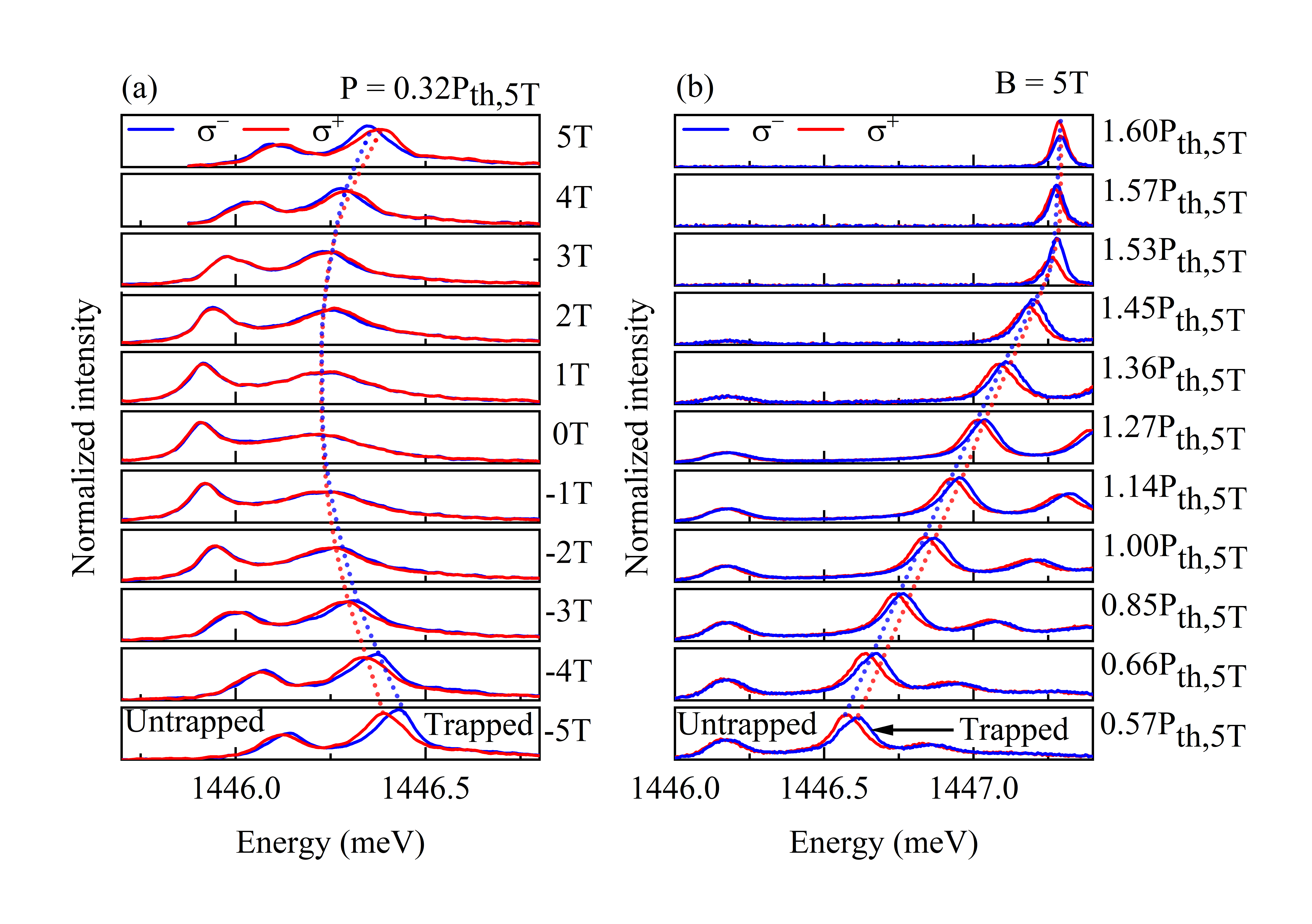}
\caption{\label{KSawicki_Fig2_magnetic_field_and_power} (a) Polarization resolved PL spectra around $k\approx0$ below condensation threshold under a changing magnetic field and constant excitation power of $P = 0.32P_{\text{th},5\text{T}}$ The diameter of the trap is $16\,\mu\mathrm{m}$. The lower and upper pair of peaks correspond to untrapped and trapped polaritons, respectively. The applied magnetic field causes a diamagnetic shift visible as a parabolic change in the emission energy of both spins. The tuning range is $\approx0.2\,\mathrm{meV}$. Another manifestation of the influence of the magnetic field is the polariton Zeeman splitting detected as an energy difference between the two opposite circularly polarized emission peaks, reaching a splitting value around $\approx24\,\mu\mathrm{eV}$ at $5\,\mathrm{T}$ for the trapped polaritons.  (b) Same as the previous panel but for varying pump powers and fixed field of $B=5$ T, showing narrowing and blueshift of the trapped spin peaks. The latter is attributed to the growing densities of interacting reservoir excitons and condensate polaritons within the trap. When the critical value ($P_{\mathrm{crit}}\approx1.5P_{\mathrm{th},5\text{T}}$) is reached, suppression of the Zeeman splitting occurs as a result of the parametric screening of the magnetic field.}
\end{figure*}

\begin{figure}[h!]
\includegraphics[width=\columnwidth]{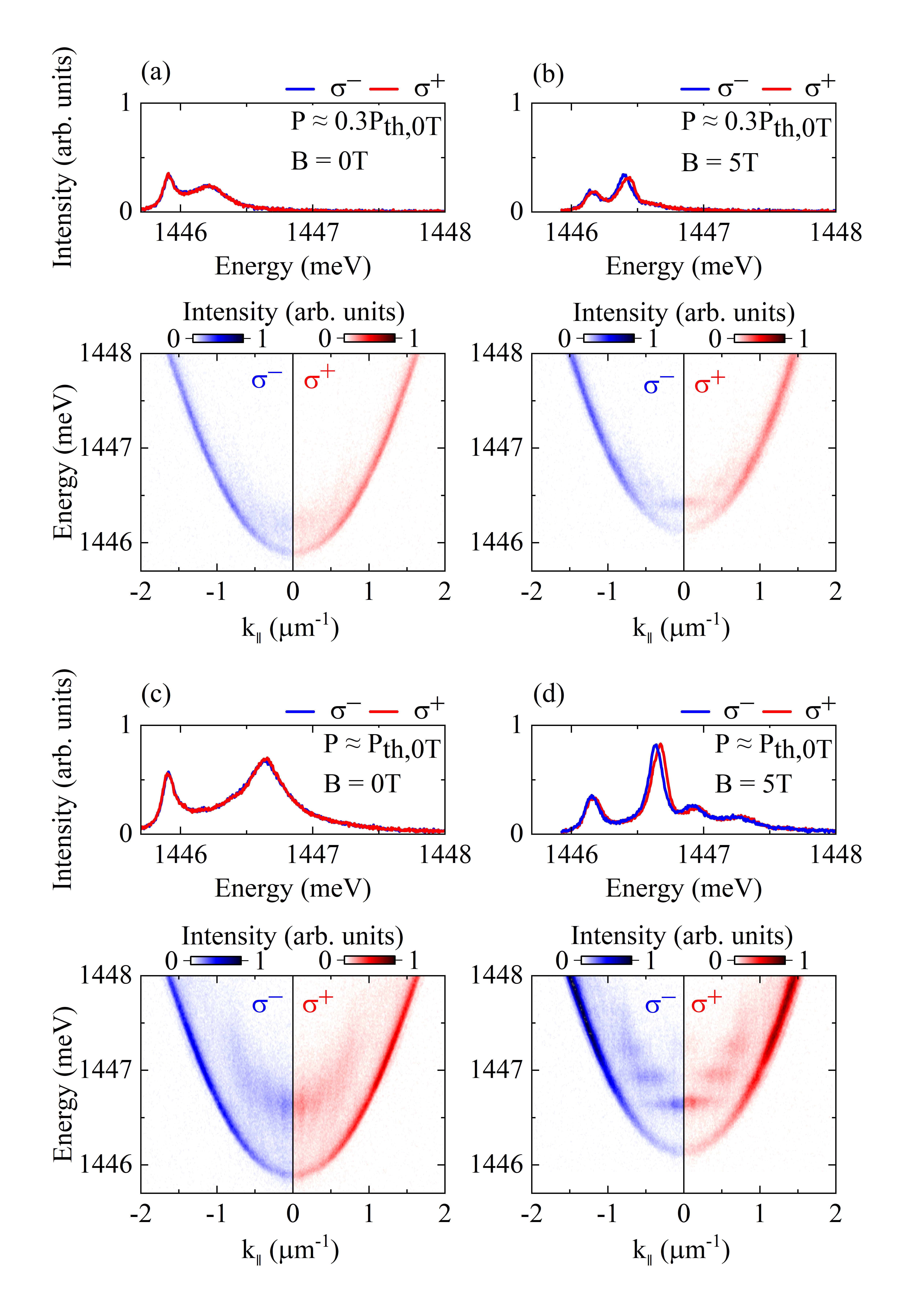}
\caption{\label{KSawicki_Fig3_kspace} Polarization resolved PL spectra (extracted around $k\approx0$) and corresponding dispersion of the polaritons measured both below and at the threshold, for both zero and finite magnetic field. The diameter of the trap is $16\,\mu\mathrm{m}$. Approaching the condensation threshold increases the density of reservoir excitons observed as a blueshift of trap ground state [compare the right peaks in (a) and (c)] with enhanced polariton intensity. When a magnetic field is applied, the condensation threshold lowers with an enhanced emission intensity of trapped polaritons [compare e.g. (c) and (d)]. The diamagnetic blueshift is also clearly visible for all peaks [compare e.g. (a) and (b)]. Notably, a ladder of trap modes starts forming with the magnetic field due to the larger exciton fraction of polaritons (i.e., the trap effectively becomes deeper).}
\end{figure}

In the absence of a magnetic field, the linearly excited sample has equal buildup of spin-up and spin-down exciton reservoir populations, resulting in the formation of a linearly polarized polariton condensate when pumped above the threshold~\cite{Kasprzak:Nature2006}. When an external magnetic field is introduced the Zeeman effect splits the energy of the spin-up and spin-down excitons, and the corresponding exciton-polariton spin states $|\psi_\pm \rangle$ as shown in Fig.~\ref{KSawicki_Fig1_scheme} (also referred by $\sigma^{\pm}$ polarized photons). The non-equilibrium spin populations of the reservoir excitons become unequal in the magnetic field because of the different spin relaxation rates~\cite{Maialle_PRB1993}. The excitons in our quantum wells are found to have a negative heavy-hole exciton $g$-factor~\cite{Snelling_PRB1992, Traynor_PRB1995} meaning that their spin ground state is antiparallel to the magnetic field. Consequently, the stimulated polaritons will preferentially condense in their spin ground state antiparallel to the field, resulting in strong circularly polarized emission of definite handedness~\cite{Kulakovskii_PRB2012, Sturm:PRB2015}.

We first study the linear regime at low pumping powers below the condensation threshold. We note that the threshold power depends on the magnetic field, and we will use the notation $P_{\text{th},B}$ where the second index refers to the field value the threshold is evaluated at. The evolution of the circularly polarized PL spectra, extracted at $k\sim0$, for consecutive values of the magnetic field is shown in Fig.~\ref{KSawicki_Fig2_magnetic_field_and_power}(a). The higher energy peak corresponds to polaritons in the ground state of the optical trap, whereas the lower energy peak is attributed to the linear PL of polaritons from the pumping area and outside the trap. Scrutinizing the higher energy peak (i.e., trapped polaritons), we observe a parabolic diamagnetic shift in both spin components~\cite{Sugawara_PRB1992} alongside the fine-structure splitting due to the Zeeman effect~\cite{Walker:PRL2011, Pietka:PRB2015, Sturm:PRB2015}.

We next fix the magnetic field at $B=5$ T in Fig.~\ref{KSawicki_Fig2_magnetic_field_and_power}(b) and increase the excitation power. We note that here each subpanel is normalized independently to keep better track of the peak location. For low pumping power values, the upper spin peaks corresponding to the trap ground state monotonically blueshift as the density of reservoir excitons and condensate polariton increases. The splitting of the two spin peaks can be clearly resolved all the way up to a critical value of $P \approx 1.5 P_{\text{th},5\text{T}}$. The lower peaks remain fixed, since they correspond to residual low-energy polaritons outside the trap. After reaching this critical value, the energy difference is eliminated completely by parametric screening of the magnetic field. This effect is known as the nonequilibrium spin-Meissner effect~\cite{Walker:PRL2011, Fischer:PRL2014, Krol:PRB2019} and, so far, has not been reported in optically trapped polariton condensates.

In Fig.~\ref{KSawicki_Fig3_kspace}, we compare the polarization- and momentum-resolved spectra for two different pump powers; $P=0.3 P_\text{th,0T}$ and $P=P_\text{th,0T}$; and two field values $B=0$ T and $B=5$ T. At low powers and zero field the polaritons are mostly unconfined and two spin degenerate dispersion branches are observed [see Fig.~\ref{KSawicki_Fig3_kspace}(a)]. Introducing a $B = 5$ T magnetic field at low powers results in diamagnetic energy shift of $0.2\,\mathrm{meV}$ and Zeeman splitting of around $24\,\mu \mathrm{eV}$ [see Fig.~\ref{KSawicki_Fig3_kspace}(b)], same as in Fig.~\ref{KSawicki_Fig2_magnetic_field_and_power}(a). Figures~\ref{KSawicki_Fig3_kspace}(c) and~\ref{KSawicki_Fig3_kspace}(d) represent the same experiment conducted at higher excitation power, which reveals a qualitative change in the dispersion when a magnetic field is applied. The magnetic field lowers the condensation threshold power~\cite{Rousset:PRB2017} in both spin components, with polaritons antiparallel to the magnetic field condensing first in the trap ground state. We note the much weaker residual emission coming from the trap excited states.

\subsection{Suppression of the Zeeman splitting}
\begin{figure}[h!]
\includegraphics[width=\columnwidth]{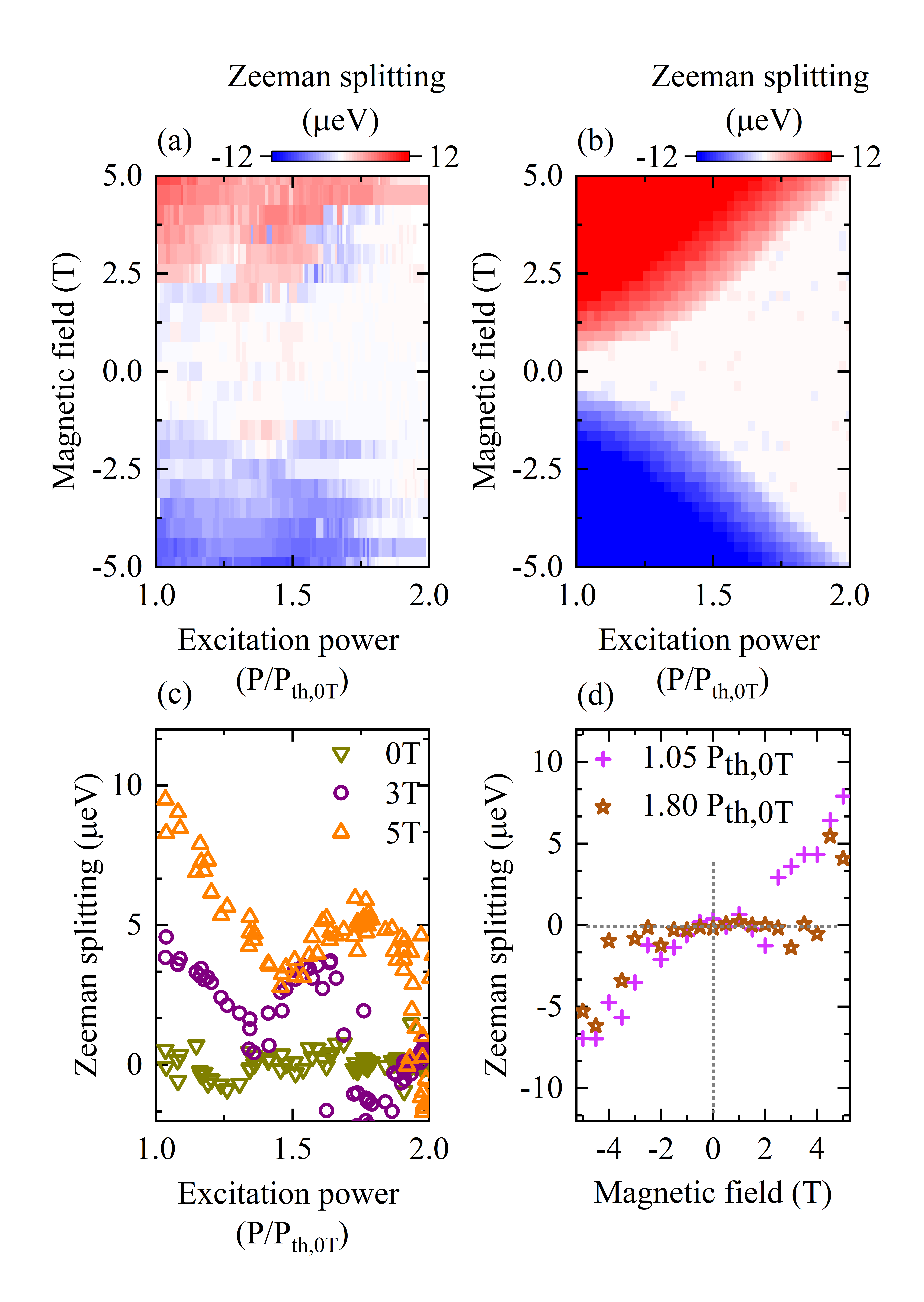}
\caption{\label{KSawicki_Fig4_suppression}(a) Zeeman splitting of the trapped polariton condensate as a function of excitation power and magnetic field. The diameter of the trap is $18\,\mu\mathrm{m}$. The splitting is clearly visible for high magnetic field strengths and low excitation power, but vanishes at certain critical boundaries in the $B$-$P$ plane (white region), becoming parametrically screened by the condensate interactions. (b) Corresponding mean field modelling (see Sec.~\ref{model}). (c) and (d) show cross-sections from panel (a) at 0T, 3T, 5T and at $P = 1.05 P_{\mathrm{th},0\mathrm{T}}$ and $P = 1.80 P_{\mathrm{th},0\mathrm{T}}$, respectively. The plots show that with increasing the excitation power, the range of magnetic field value in which Zeeman splitting is completely suppressed expands.}
\end{figure}
We now perform a systematic scan over both the pump power and the magnetic field and scrutinize the changes in the position of the polarization resolved spectral peaks in the trapped condensate. Figure~\ref{KSawicki_Fig4_suppression}(a) shows the average Zeeman splitting of the condensate as a function of excitation power and magnetic field (i.e., each pixel represents the average over many condensate realizations). Two regimes of opposite splitting can be clearly seen, separated by a region of negligible splitting. In the latter, the condensate components become equal in terms of population and emission energy, resulting in a strongly linearly polarized emission from the condensate. The non-gradual change in the splitting implies an abrupt change in the condensate dynamics, wherein the Zeeman splitting becomes fully suppressed. The experimental result are reproduced through numerical simulations using a generalized Gross-Pitaevskii equation describing a spinor condensate order parameter coupled to excitonic reservoirs [see Fig.~\ref{KSawicki_Fig4_suppression}(b) and  Sec.~\ref{model}].

We select three horizontal and two vertical cross-sections from Fig.~\ref{KSawicki_Fig4_suppression}(a) and plot them in Fig.~\ref{KSawicki_Fig4_suppression}(c) and Fig.~\ref{KSawicki_Fig4_suppression}(d) for clarity. Fig.~\ref{KSawicki_Fig4_suppression}(c) shows a non-monotonic behaviour in the Zeeman splitting as a function of power, whose origin may be related to a pump-induced ellipticity with increasing the magnetic field due to birefringence of the pumping optics.

\subsection{Magnetically induced spin inversion}

\begin{figure}[t!]
\includegraphics[width=9cm]{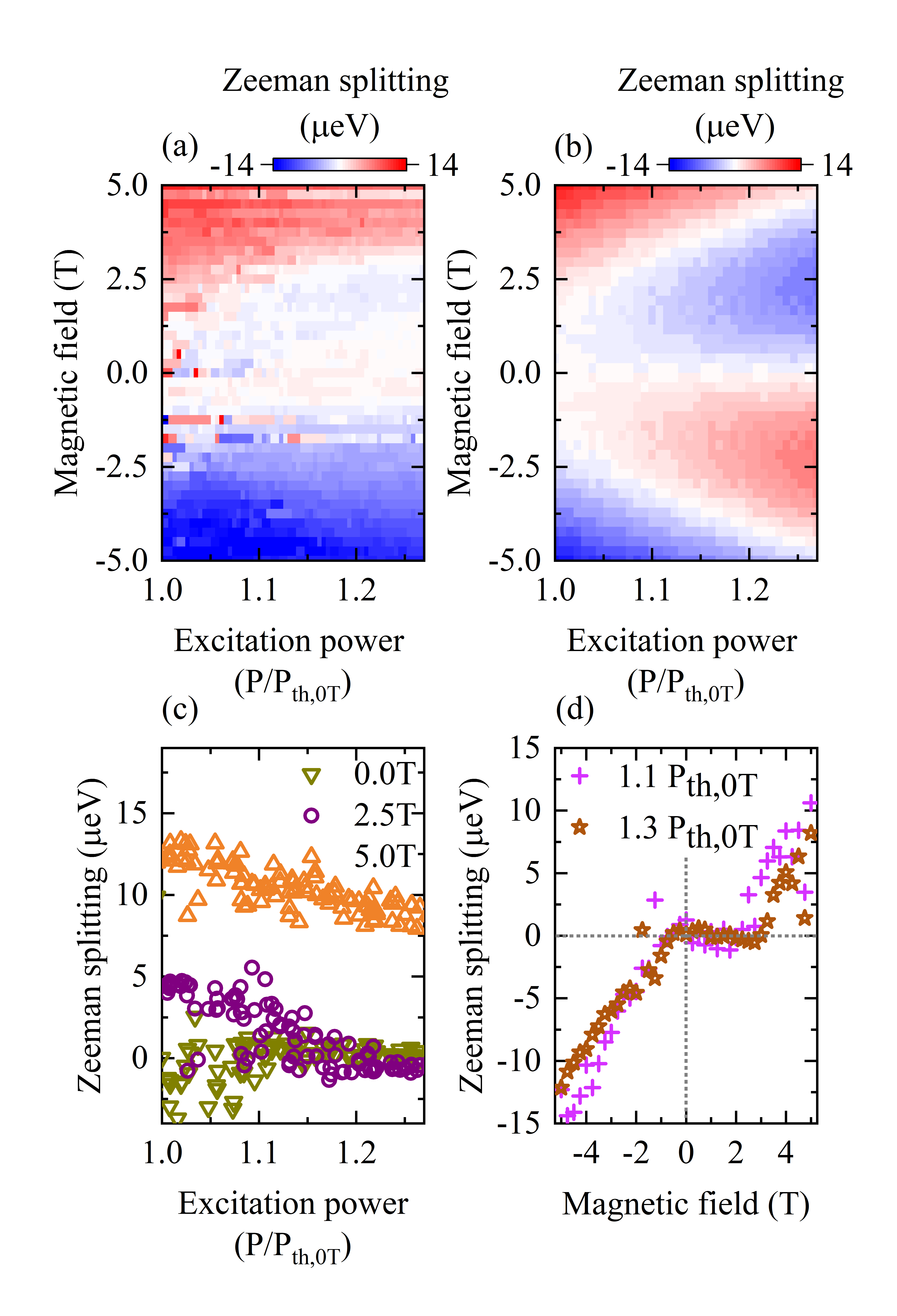}
\caption{\label{KSawicki_Fig5_inversion} Power induced inversion of the Zeeman splitting. The diameter of the trap is here $27.2\,\mu\mathrm{m}$. (a) Energy splitting as a function of the magnetic field and excitation power. The emission energy of both circular polarization components is measured simultaneously. The plot presents the partial suppression of Zeeman splitting and the reversal of sign after exceeding the critical value of the excitation power. (b) Corresponding mean field modelling (see Sec.~\ref{model}). (c) and (d) show horizontal and vertical cross-sections of the data in panel (a), respectively.}
\end{figure}

The Zeeman suppression presented in the previous section occurs for relatively small optical traps (around $\lesssim 20$ $\mu$m). In this section, we describe a qualitatively different regime by increasing the size of the optical trap, consequently reducing the confinement of the condensate. In particular, the Zeeman suppression is now accompanied by a regime of power induced inverted Zeeman splitting,  as shown in Fig.~\ref{KSawicki_Fig5_inversion}(a). A similar inversion was reported recently in optically trapped condensates where an elliptically polarized excitation beam replaced the role of a real magnetic field~\cite{delValle-InclanRedondo:PRB2019}. Also, an inversion of the Zeeman splitting was reported for the excited
quantum-confined states in wider InGaAs quantum wells~\cite{Grigoryev:PRB2016} and quantum wires under high magnetic field~\cite{Hammersberg:JJAP1997}. Mean field simulations taking into account the increased size of the trap also reproduce the experimental observation as shown in Fig.~\ref{KSawicki_Fig5_inversion}(b) [see Sec.~\ref{sec.GPE} for details]. We select three horizontal and two vertical cross-sections from Fig.~\ref{KSawicki_Fig5_inversion}(a) and plot them in Fig.~\ref{KSawicki_Fig5_inversion}(c) and Fig.~\ref{KSawicki_Fig5_inversion}(d) for
clarity. We note that the slight asymmetry of the observed splitting about $B=0$ in Fig.~\ref{KSawicki_Fig5_inversion}(a) is due to the small polarization ellipticity in the pump beam.

\begin{figure}[t]
\includegraphics[width=9cm]{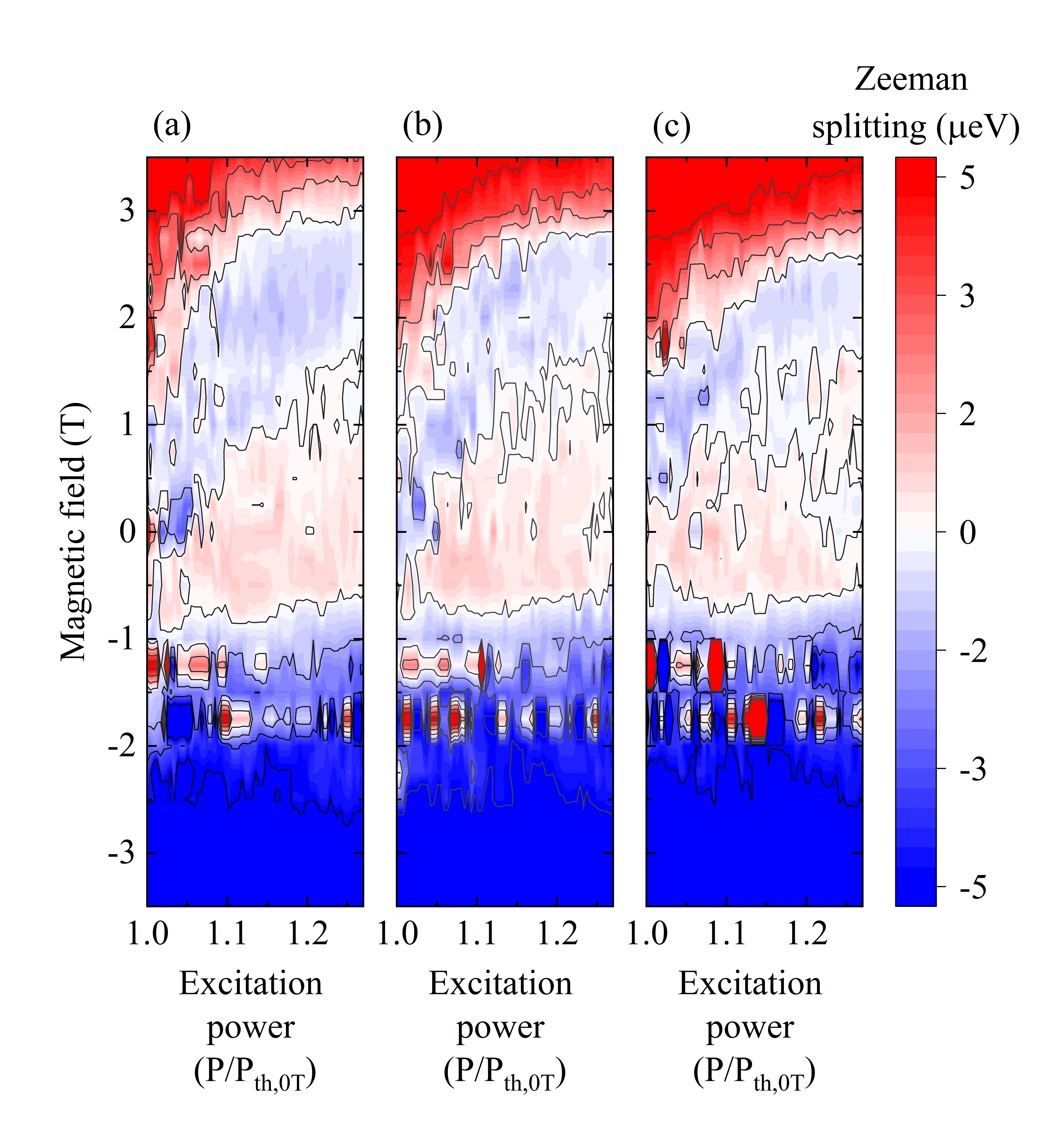}
\caption{\label{KSawicki_Fig6_inversion} The map of Zeeman splitting for different sizes of the trap: (a) 27.2 $\mu\mathrm{m}$ (b) 26.4 $\mu\mathrm{m}$ (c) 25.6 $\mu\mathrm{m}$. In contrast to the suppression regime, where after exceeding the critical value of power, the Zeeman splitting is fully suppressed, in the inversion regime, exceeding the critical value results in the reversal of the Zeeman splitting sign. The value of the critical power that triggers the inversion depends on the applied magnetic field and the size of the optical trap.}
\end{figure}

In Fig.~\ref{KSawicki_Fig6_inversion} we compare the Zeeman splitting inversion for three different sizes of the trap: (a) $27.2\,\mu\mathrm{m}$ (b) $26.4\,\mu\mathrm{m}$ (c) $25.\,\mu\mathrm{m}$. We observe a decrease in the inversion region with decreasing trap size. In addition, the boundary between regions with positive and negative Zeeman splitting is noticeably blurred and the area in which Zeeman splitting is completely suppressed noticeably increases. This can be attributed to the beginning of the transition from the inversion regime to the suppression regime. As we mentioned before, the asymmetry of the splitting pattern about $B=0$ in Fig.~\ref{KSawicki_Fig6_inversion} is due to the small polarization ellipticity in the pump beam.

\section{\label{model} Theoretical model}
In this section, we describe how the magnetic field changes the exciton spin levels and how these effects enter into the polariton states through strong light-matter coupling. We then move onto a many-body mean field treatment where we introduce a zero-dimensional generalized Gross-Pitaevskii equation describing the condensate spinor order parameter coupled to excitonic reservoirs~\cite{Ohadi:PRX2015}. Numerically solving the equations of motion allows us to produce qualitatively the experimental observations in Fig.~\ref{KSawicki_Fig4_suppression}(b) and Fig.~\ref{KSawicki_Fig5_inversion}(b).

\subsection{Single spinor polaritons in a magnetic field}
The energy of excitons in III-V semiconductor is modified by the magnetic field $B$ in the following way,
\begin{equation}
E_{\mathrm{X},\pm} = E_{\mathrm{X},0} \mp g_\text{X}\mu_{\mathrm{B}} B + \gamma_{\mathrm{dia}} B^2.
\label{eq1_Zeeman_and_diamagneitic_shift}
\end{equation}
Here, $E_{\mathrm{X},0}$ is the bare exciton energy, $\mu_{\mathrm{B}}$ is the Bohr magneton, $g_\text{X}$ is the exciton Landé $g$-factor, and $\gamma_{\mathrm{dia}}$ quantifies the exciton diamagnetic shift. Both coefficients determining the magnetic response of the exciton level can be estimated from fitting to experiment giving $g_\text{X}=-0.364$, in good agreement for 6 nm In$_{0.08}$Ga$_{0.92}$As QWs~\cite{Traynor_PRB1995}, and $\gamma_{\mathrm{dia}}=0.117\,\mathrm{meV}\,\mathrm{T}^{-2}$ (see Appendix~\ref{APPENDIX_A_g_and_gamma } for details).

The energies of the spin-up and spin-down lower-branch polaritons follow directly from a standard coupled oscillator model~\cite{Carusotto:RMP2013}:
\begin{equation}
E_\pm = \frac{E_{\mathrm{ph}} + E_{\mathrm{X},\pm}}{2} - \frac{1}{2}\sqrt{4\Omega^2 + \Delta_{\pm}^2}.
\label{eq1_energy_of_polaritons}
\end{equation}
Here, $\Omega$ is the Rabi energy (i.e., light-matter coupling strength), $E_{\mathrm{ph}}$ is the cavity photon energy, and $\Delta_{\pm}=E_{\mathrm{ph}} - E_{\mathrm{X},\pm}$ is the detuning. In the absence of any pumping, the polariton Zeeman splitting is simply,
\begin{equation}
E_{\mathrm{ZS}}^{(0)} =-g_{\mathrm{X}}\mu_{\mathrm{B}}B +\frac{1}{2}\left[\sqrt{\Delta_{-}^2+4\Omega^2}-\sqrt{\Delta_{+}^2+4\Omega^2}\right].
\label{eq1_zeeman_splitting}
\end{equation}
The squared brackets are related to the matter content of the polaritons which is given by the exciton Hopfield coefficient,
\begin{equation} \label{hopfield}
    |X_\pm|^2 = \frac{1}{2}\left( 1 + \frac{\Delta_\pm}{\sqrt{ \Delta_\pm^2 +  4\Omega^2}} \right).
\end{equation}
Moreover, under the influence of the magnetic field, the wave function of the exciton decreases, which results in an increase of the exciton oscillator strength and, consequently, Rabi splitting \cite{Fisher:PRB1996, Berger:PRB1996, Stepnicki:PRB2015},
\begin{equation}
\begin{split}
\Omega &= \frac{\Omega_{0}}{\sqrt{2}}\left[\sqrt{1+\frac{3}{2}\left(\frac{e^2a_{0}^4B^2}{\hbar^2}\right)}+1\right]^{\frac{1}{2}}.
\end{split}
\end{equation}
Here, $\Omega_{0}$ and $a_{0}$ are bare Rabi energy and Bohr radius of exciton, respectively. The above expressions determine the Zeeman splitting of lower-branch polariton in the single-particle limit. 

\subsection{Generalized Gross-Pitaevskii model} \label{sec.GPE}
To understand the effect of the condensate and background excitons on the polariton spin energy levels we start by defining the condensate spinor order parameter as $\Psi \equiv (\psi_{+}, \psi_{-})^\text{T}$, describing the phase and population $|\psi_\pm|^2 \equiv \rho_\pm$ of each spin component in the optical trap ground state. The condensate state is often conveniently described in terms of it pseudospin,
\begin{equation}
    \mathbf{S} \equiv \frac{\langle \Psi | \hat{\boldsymbol{\sigma}}| \Psi \rangle}{\langle \Psi | \Psi \rangle},
\end{equation}
where $\hat{\boldsymbol{\sigma}}$ is a vector of Pauli matrices. The components of the condensate pseudospin are proportional to the Stokes parameters of the emitted cavity light, permitting full spin characterization through standard polarimetry measurements~\cite{Gnusov_PRB2020}.

The corresponding zero-dimensional generalized Gross-Pitaevskii equation coupled to the rate equation describing a hot background exciton reservoir $n_\pm$ can be written~\cite{Ohadi:PRX2015}:
\begin{eqnarray}
i\hbar\frac{d {\psi}_{\pm}}{d t} &=& \left[ E_\pm + \frac{i\hbar}{2}(R n_\pm - \gamma) + \alpha_{\pm}|\psi_\pm|^2 \right. \nonumber
\\
&&\left. + G_{\pm}\left(n_\pm + \frac{\mathcal{P}_\pm}{W} \right) \right] \psi_{\pm} + \frac{E_{\mathrm{XY}}}{2} \psi_\mp,
\label{Gross_Pitaevskii_equation}
\\
\frac{d n_\pm}{d t} &=& - (\Gamma + \Gamma_{s,\pm} + R|\psi_\pm|^2)n_\pm \nonumber \\
&& + \Gamma_{s,\mp}n_\mp + \mathcal{P}_\pm,
\label{reservoir_rate_equation}
\end{eqnarray}
where $\gamma^{-1}$ is the polariton lifetime and $R$ quantifies the scattering rate of reservoir excitons into the condensate. We will neglect opposite-spin polariton interactions, since they are much weaker than same-spin interactions~\cite{Ciuti:PRB1998, Bieganska:PRL2021}, and only include same-spin polariton-reservoir $G_{\pm} = 2u_{\mathrm{X}}{|X_{\pm}|^2}$ and polariton-polariton $\alpha_{\pm} = \xi u_{\mathrm{X}}|X_{\pm}|^4$ interactions. Here, $u_{\mathrm{X}}$ stands for the exciton-exciton Coulomb interaction strength normalized over the number of quantum wells. $\Gamma$ is the reservoir decay rate, $\Gamma_{\mathrm{s},\pm}$ describes the rate of spin relaxation for each spin component. We also account for finite linear polarization splitting by the parameter $E_{\mathrm{XY}}$ coming from small random birefringence due to sample strain and disorder~\cite{Kasprzak:Nature2006, Ohadi:PRX2015, Gnusov_PRB2020}. 

Here we have introduced a dimensionless parameter $\xi$ describing the ratio of the two overlap integrals,
\begin{equation} \label{eq.overlap}
    \xi_\sigma = \frac{\int | \langle \mathbf{r} | \psi_\sigma \rangle|^4 \, d\mathbf{r}}{\int |\langle \mathbf{r} | \psi_\sigma \rangle|^2 n_\sigma(\mathbf{r}) \, d\mathbf{r}}.
\end{equation}
Here, $n_\pm(\mathbf{r}) \propto P(\mathbf{r})$ is the exciton reservoir density in real space, which is proportional to the nonresonant pump profile. For simplicity we will assume that the spatial profiles for both spins are similar so that $\xi_\pm \simeq \xi$. Equation~\eqref{eq.overlap} comes from projecting the the two-dimensional Gross-Pitaevskii equation into a zero-dimensional equation. Because the amplitudes of the reservoir and the condensate can be scaled arbitrarily by adjusting $R$ it is safe to fix all overlap-dependent parameters and allow only $\xi$ to vary when the trap size is changed. Larger traps result in higher pump threshold and larger condensate population~\cite{Gnusov_PRB2020} implying that $\xi$ scales in proportion with trap size.

The parameter $W>0$ quantifies the conversion of dark and high-momentum inactive excitons $\mathcal{P}_\pm$ into the active "bottleneck" exciton reservoir $n_\pm$ which sustains the condensate. Under continuous-wave excitation we can approximate the steady state of the inactive reservoir as~\cite{Pickup:PRB2021}:
\begin{equation}
\binom{\mathcal{P}_+}{\mathcal{P}_-} = \frac{P}{W + 2\Gamma_{s} } \binom{W \cos^2(\theta) + \Gamma_{s,-}}{W \sin^2(\theta) + \Gamma_{s,+}}
\label{expression_of_power}
\end{equation}
Here, $P$ denotes the power of the nonresonant pump, and $\theta$ is analogous to the quarter wave plate angle for the incident excitation, which defines the ellipticity of the excitation. We will consider a linearly polarized excitation with $\theta=\pi/4$ throughout the study.

We account for the influence of the magnetic field on the exciton spin relaxation~\cite{Crooker:PRB1997} using polynomial regression to fit the different spin-relaxation rates to the behaviour of the exciton reservoir degree of circular polarization below threshold. The general form of the spin relaxation parameters can be written,
\begin{equation}
\Gamma_{s,\pm} = \Gamma_s \pm \eta(B),
\label{Gamma_S}
\end{equation}
where the details of determining the function $\eta$ are presented in Appendix~\ref{APPENDIX_B_eta}.

Numerically solving Eqs.~\eqref{Gross_Pitaevskii_equation} and~\eqref{reservoir_rate_equation} in time we are able to qualitatively reproduce the experimental observations. For small traps in Fig.~\ref{KSawicki_Fig4_suppression}(b) we show the energy splitting coming from solving the generalized Gross-Pitaevskii equation~\eqref{Gross_Pitaevskii_equation} from random initial conditions for $\xi = 20$ and $E_{XY} = 0.6$ ns$^{-1}$. All other parameters are kept fixed and can be found in Appendix~\ref{APPENDIX_C_parameters}.

For a given external magnetic field in Fig.~\ref{KSawicki_Fig4_suppression} $\mathbf{B}_\text{ext}$, the suppression of the fine structure splitting happens above a certain critical pump power  $P > P_{\mathrm{crit}}$ due to the spin-anisotropic interactions of polaritons~\cite{Ciuti:PRB1998, Martin_PRL2002}. When the external magnetic field splits the spin levels (see Fig.~\ref{KSawicki_Fig1_scheme}) condensate polaritons start accumulating in the spin ground state. As the condensate becomes more circularly polarized, it produces its own interaction-induced magnetic field written,
\begin{equation}\label{eq.Bint}
\mathbf{B}_\text{int} \propto (\alpha_+ \rho_+ - \alpha_- \rho_- + G_+ n_+ - G_-n_-)  \mathbf{\hat{z}}.
\end{equation}
With increasing power, the condensate and reservoir spin population imbalance increase, eventually generating a strong enough effective field which counter the real field until,
\begin{equation} \label{eq.netfield}
|\mathbf{B}_\text{ext} + \mathbf{B}_\text{int}| \lesssim |\mathbf{B}_\parallel| = E_{XY} /\mu_B.
\end{equation}
When the total out-of-plane field is weak compared to the effective in-plane field, an elliptically polarized fixed point attractor forms in the phase space of the condensate which ``pins'' the pseudospin~\cite{Read_PRB2009, Gnusov_PRB2020} and suppresses the energy difference of emitted $\sigma^\pm$ photons. This corresponds to the white region in Fig.~\ref{KSawicki_Fig4_suppression}(a) and~\ref{KSawicki_Fig4_suppression}(b). In the blue and red regions the condensate converges into a limit cycle with the pseudospin undergoing a tilted precession on the Poincaré sphere~\cite{Gnusov_PRB2020} at a rate given by the energy splitting. It is important to note that without $E_{XY}$ no stable fixed point attractors are found in simulation for low magnetic fields $B$ and the sharp boundary of the white region in Fig.~\ref{KSawicki_Fig4_suppression}(b) is absent.

To reproduce the Zeeman inversion shown in Fig.~\ref{KSawicki_Fig5_inversion} for a larger traps we need to increase the interaction-induced magnetic field for the corresponding power range. We observe that the larger trap has a higher threshold and higher condensate population relative to the condensation point~\cite{Gnusov_PRB2020}. This implies that $\xi$ must be increasing with trap size. For this purpose we increase the condensate blueshift term to $\xi = 47$, and also set $E_{XY}=0$ although this is not strictly necessary (see Appendix~\ref{AppendixE}). The resulting energy splitting coming from mean field simulations is presented in Fig.~\ref{KSawicki_Fig5_inversion}(b) showing good qualitative agreement with experiment. We note that, the local birefringence $E_{XY}$ can be different between the two experiments in Fig.~\ref{KSawicki_Fig4_suppression} and~\ref{KSawicki_Fig5_inversion} because two different sample location were used without affecting the exciton-
photon detuning.

Therefore, by adjusting only two parameters, $\xi$ and $E_{XY}$ we are able to reproduce both the reported screening (pseudospin pinning) and the inversion (pseudospin precession reversal) effects of our experiment. A more detailed comparison of how these parameters change the energy splitting of the spins is given in Appendix~\ref{AppendixE}.

\subsection{Spinor condensate solution}
Here we derive an analytical expression for the Zeeman splitting in Eq.~\eqref{Zeeman_splitting_analytical_formula} complementing our numerical simulations.

Above the condensation threshold, where one or both condensate spin populations $\rho_\pm$ are finite and positive, the Zeeman splitting~\eqref{eq1_zeeman_splitting} becomes modified due to reservoir saturation (i.e., gain clamping) and interactions from the condensate. The below threshold solution is given in Appendix~\ref{AppendixD}. Focusing on the optically isotropic cavity where $E_{XY} = 0$ and assuming that the condensate spin populations have reached a steady state $\dot{\rho}_\pm = 0$ for given pump power and magnetic field, we can write a general ansatz,
\begin{equation}
\Psi = \binom{\sqrt{\rho_{+}}e^{-iE_{\mathrm{c},+}t/\hbar}}{\sqrt{\rho_{-}}e^{-iE_{\mathrm{c},-}t/\hbar}},
\label{steady_state_psi}
\end{equation}
where
\begin{eqnarray}E_{c,\pm} =  E_\pm + \alpha_{\pm} \rho_\pm + G_{\pm}\left(n_\pm + \frac{\mathcal{P}_\pm}{W} \right) .
\end{eqnarray}
This solution describes a precessing pseudospin, similar to the self-induced Larmor precessing condensate pseuodspins reported for elliptically polarized pumps~\cite{Baryshev_PRL2022, Sigurdsson_PRL2022}.

Substituting Eq.~\eqref{steady_state_psi} into Eq.~\eqref{Gross_Pitaevskii_equation}, we arrive at
%
%
\begin{equation} \label{eq.cond_sol}
\begin{split}
\rho_\pm & = \left(\frac{\mathcal{P}_\pm}{\gamma} + \frac{ \frac{\Gamma_{s,\mp}\mathcal{P}_\mp}\gamma - \frac{\Gamma(\Gamma + 2 \Gamma_s)}{R}}{\Gamma + \Gamma_{s,\mp}} \right)  \\ 
& \times H[\mathcal{P}_\mp - \mathcal{P}_{\text{th},\mp}] H[\mathcal{P}_{\text{th},\pm} - \mathcal{P}_\pm ] \\
& + \left(\frac{\mathcal{P}_\pm}{\gamma} - \frac{ \Gamma + \Gamma_{s,\pm} - \Gamma_{s,\mp}}{R} \right) H[\mathcal{P}_\pm - \mathcal{P}_{\text{th},\pm}],
\end{split}
\end{equation}
where $H[\dots]$ is the Heaviside function and $\mathcal{P}_{\text{th},\pm}$ denotes the condensation threshold for each spin component. The first bracket denotes a fully circularly polarized condensate where only one spin-component, with the lower threshold, has condensed. The second bracket describes an elliptically polarized solution when both components are above threshold. 

\begin{figure}
    \centering
    \includegraphics[width=\linewidth]{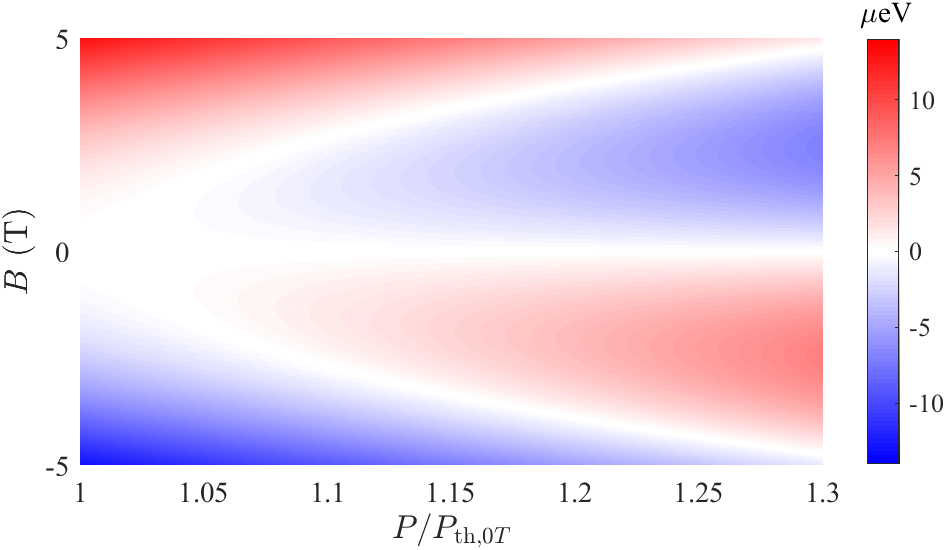}
    \caption{\label{KSawicki_Fig7_analytical} Inversion of the Zeeman splitting from the analytical solution~\eqref{Zeeman_splitting_analytical_formula} using same parameters as in Fig.~\ref{KSawicki_Fig5_inversion}(b).}
\end{figure}

The solution for the reservoir is segmented into three terms,
\begin{align} \label{eq.res}
    n_\pm &= \frac{(\Gamma + \Gamma_{s,\mp})\mathcal{P}_\pm + \Gamma_{s,\mp} \mathcal{P}_\mp }{\Gamma(\Gamma + 2 \Gamma_s)}H[\mathcal{P}_{\text{th},\pm} - \mathcal{P}_\pm ] \\
    & + \frac{\gamma \Gamma_{s,\mp} /R + \mathcal{P}_\pm}{\Gamma + \Gamma_{s,\pm}}  H[\mathcal{P}_\mp - \mathcal{P}_{\text{th},\mp}] H[\mathcal{P}_{\text{th},\pm} - \mathcal{P}_\pm ]  \nonumber \\
    & + \frac{\gamma}{R} H[\mathcal{P}_\pm - \mathcal{P}_{\text{th},\pm}]  \nonumber.
\end{align}
Here, the first term describes the reservoir occupation when both condensate spin components are below threshold (no condensate); the second term when one component is above threshold (circularly polarized condensate); the third term when both components are above threshold (elliptically polarized condensate). Note that 
at zero magnetic field the two thresholds coincide and can be written simply, 
\begin{equation}
P_{\mathrm{th},0\mathrm{T}} = \frac{2 \gamma \Gamma}{R}.
\label{threshold_power_0T}
\end{equation}
The complete expression for the Zeeman splitting is,
\begin{eqnarray}
E_{\mathrm{ZS}} &=& E_{\mathrm{ZS}}^{(0)} + G_{+}\left(n_+ + \frac{\mathcal{P}_{+}}{W}\right)-G_{-}\left(n_- + \frac{\mathcal{P}_{-}}{W} \right) \nonumber
\\
&&  + \alpha_{+}\rho_+ - \alpha_{-}\rho_-.
\label{Zeeman_splitting_analytical_formula}
\end{eqnarray}
In Fig.~\ref{KSawicki_Fig7_analytical} we plot Eq.~\eqref{Zeeman_splitting_analytical_formula} for the same set of parameters as used in the numerical simulation presented in Fig.~\ref{KSawicki_Fig5_inversion}, showing excellent agreement between numerics and the derived Zeeman splitting using~\eqref{steady_state_psi}.

\section{Conclusions}
In conclusion, we have demonstrated the appearance of Zeeman splitting of an optically trapped polariton condensate subjected to an external magnetic field along the optical axis of a planar microcavity embedded with several pairs of InGaAs quantum wells. We explain the conditions needed to obtain two operation regimes, where we observe two qualitatively different spin-related effects: 1) the full parametric screening of external magnetic field and the suppression of the Zeeman splitting; 2) the inversion of the spin-population and sign reversal of the Zeeman splitting. We develop a mean field model based on a zero-dimensional generalized Gross-Pitaevskii equation coupled to a rate equation for the exciton reservoir, which qualitatively captures the observed effects.

Optical trapping of polariton condensate offers a powerful tool for magneto-optical studies of microcavities, which has not been explored so far. The advantage of the technique used in this work is its tunability and reconfigurability. The efficient manipulating of spins with a magnetic field in systems with arbitrarily arranged pump geometry opens up new possibilities for a wide class of magnetically controlled polariton devices. The single optical trap system presented in this study is the building block, which can be adapted to complex polariton systems, such as polariton lattices or coupled trapped condensates. As an example, a natural next step could be investigating magnetic manipulation of synchronization and emission properties between spatially coupled~\cite{Harrison:PRB2020, Gnusov:PRA2021} or time modulated~\cite{Gnusov:arxiv2023} optical traps.

\appendix

\section{\label{APPENDIX_A_g_and_gamma } Estimation of the excitonic Landé $g$-factor and diamagnetic coefficient}
\begin{figure}[t!]
\includegraphics[width=\columnwidth]{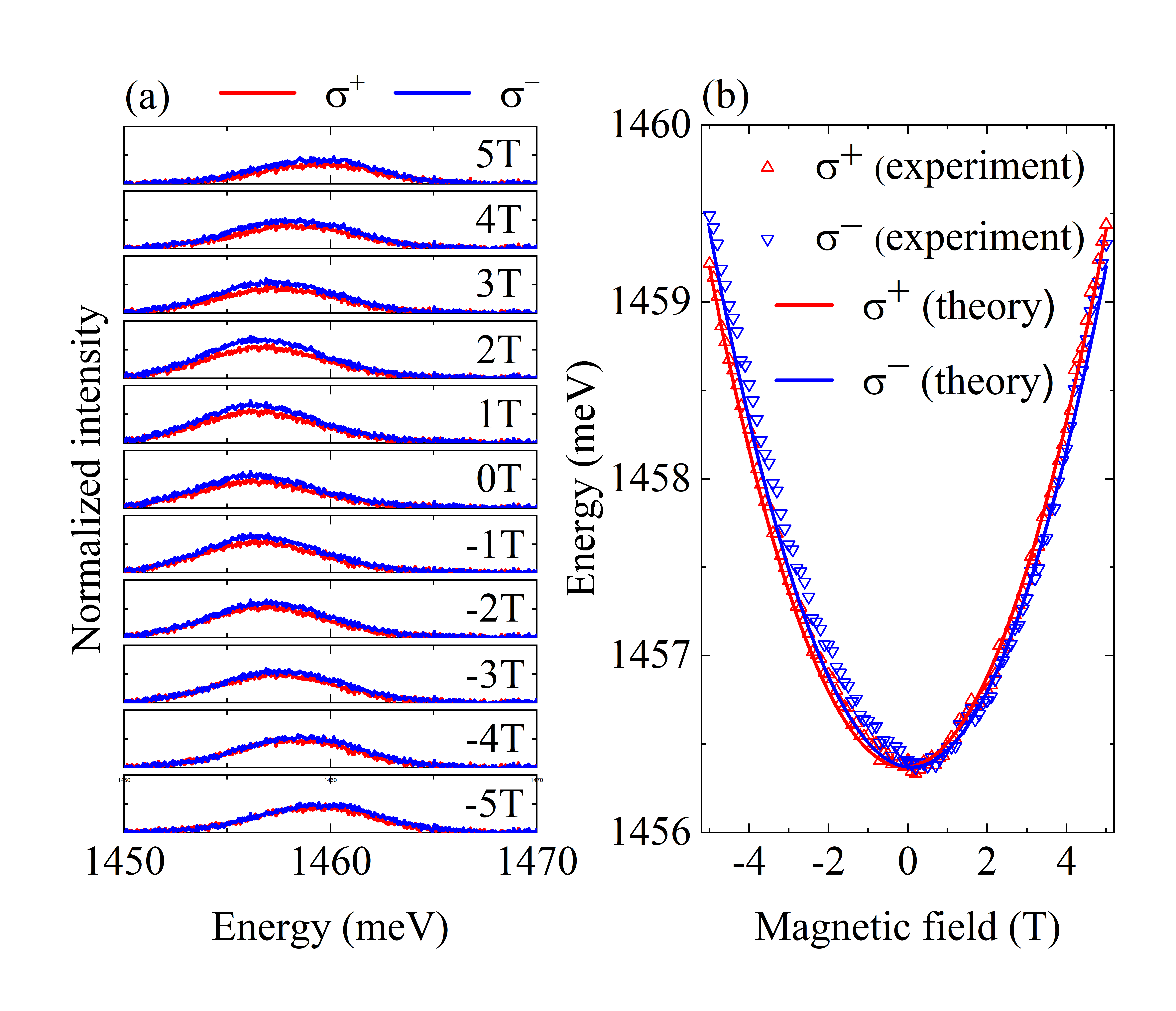}
\caption{\label{KSawicki_Fig8_exciton} (a) PL spectra of the emission from excitonic level at an increasing magnetic field. With the magnetic field, the Zeeman splitting value increases, reaching $\sim0.21\,\mathrm{meV}$ at 5T. (b) Shift of exciton energies with applied magnetic field for $\sigma^+$ (red) and $\sigma^-$ (blue) polarization of detection; solid lines show fits using  Eq.~\eqref{KSawicki_Fig8_exciton}.
}
\end{figure}

Here we present the data used to determine $g_\text{X}$, the exciton Landé $g$-factor, and $\gamma_{\mathrm{dia}}$, the exciton diamagnetic shift parameter, in Eq.~\eqref{eq1_Zeeman_and_diamagneitic_shift}. Figure~\ref{KSawicki_Fig8_exciton}(a) shows polarization resolved exciton line (below threshold) and~\ref{KSawicki_Fig8_exciton}(b) shows the peak value of each spectral component as a function of magnetic field. Solid lines are the fit using Eq.~\eqref{eq1_Zeeman_and_diamagneitic_shift} giving $g_\text{X}=-0.364$, in good agreement for 6 nm In$_{0.08}$Ga$_{0.92}$As QWs~\cite{Traynor_PRB1995}, and $\gamma_{\mathrm{dia}}=0.117\,\mathrm{meV}\,\mathrm{T}^{-2}$.

\section{\label{APPENDIX_B_eta} Dependence of reservoir spin relaxation on magnetic field}
In this section, we determine the function $\eta(B)$ in Eq.~\eqref{Gamma_S} which describes how the magnetic field modifies the exciton spin relaxation rate $\Gamma_s$. This effect leads to different steady state populations of the reservoir exciton spins under linearly polarized pumping $\theta=\pi/4$ below condensation threshold ($|\psi_\pm|^2=0$). Consequently, it contributes to the different blueshifts experienced by the trapped polariton spins when a magnetic field is present [see the term proportional to $G_\pm$ in Eq.~\eqref{Gross_Pitaevskii_equation}]. 

In order to estimate the modified spin relaxation rates of the excitons we scrutinize the degree of circular polarization (DCP) of the emitted photons below threshold. Assuming that most of the below-threshold emission is coming from reservoir excitons in the bottleneck region we can define the below-threshold DCP in terms of the reservoir densities $n_\pm$ in a standard way,
\begin{equation}
\mathrm{DCP} = \frac{n_{+}-n_{-}}{n_{+}+n_{-}}.
\end{equation}

\begin{figure}[t!]
\includegraphics[width=\columnwidth]{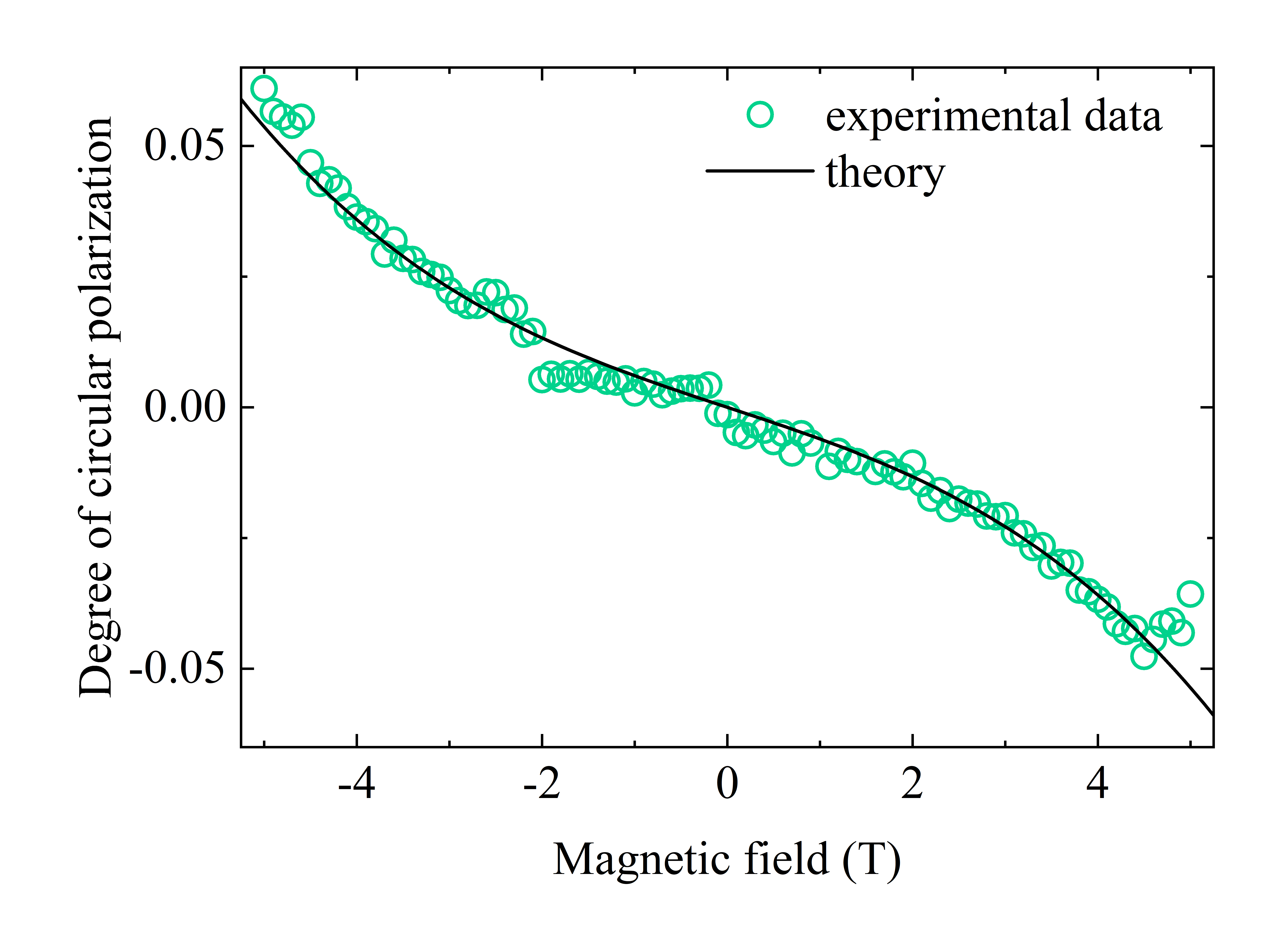}
\caption{\label{KSawicki_Fig9_DOP} Degree of circular polarization of the cavity photoluminescence below the threshold as function of magnetic field.}
\end{figure}

The steady states of $n_\pm$ below the threshold can be trivially obtained from Eq.~\eqref{reservoir_rate_equation}. We can fit the above equation to our experimental data (see Fig.~\ref{KSawicki_Fig9_DOP}) using a simple polynomial model for the change in spin relaxation,
\begin{eqnarray}
\eta(B) = \eta_{1} B + \eta_{3} B^3,
\label{eta_function}
\end{eqnarray}
where $\eta_1$ and $\eta_3$ are fitting parameters. For the purpose of this study, we found that sufficiently good agreement between experiment and theory was found by only considering the lowest order $\eta_1$ contribution.
\begin{figure*}
\includegraphics[width=0.8\linewidth]{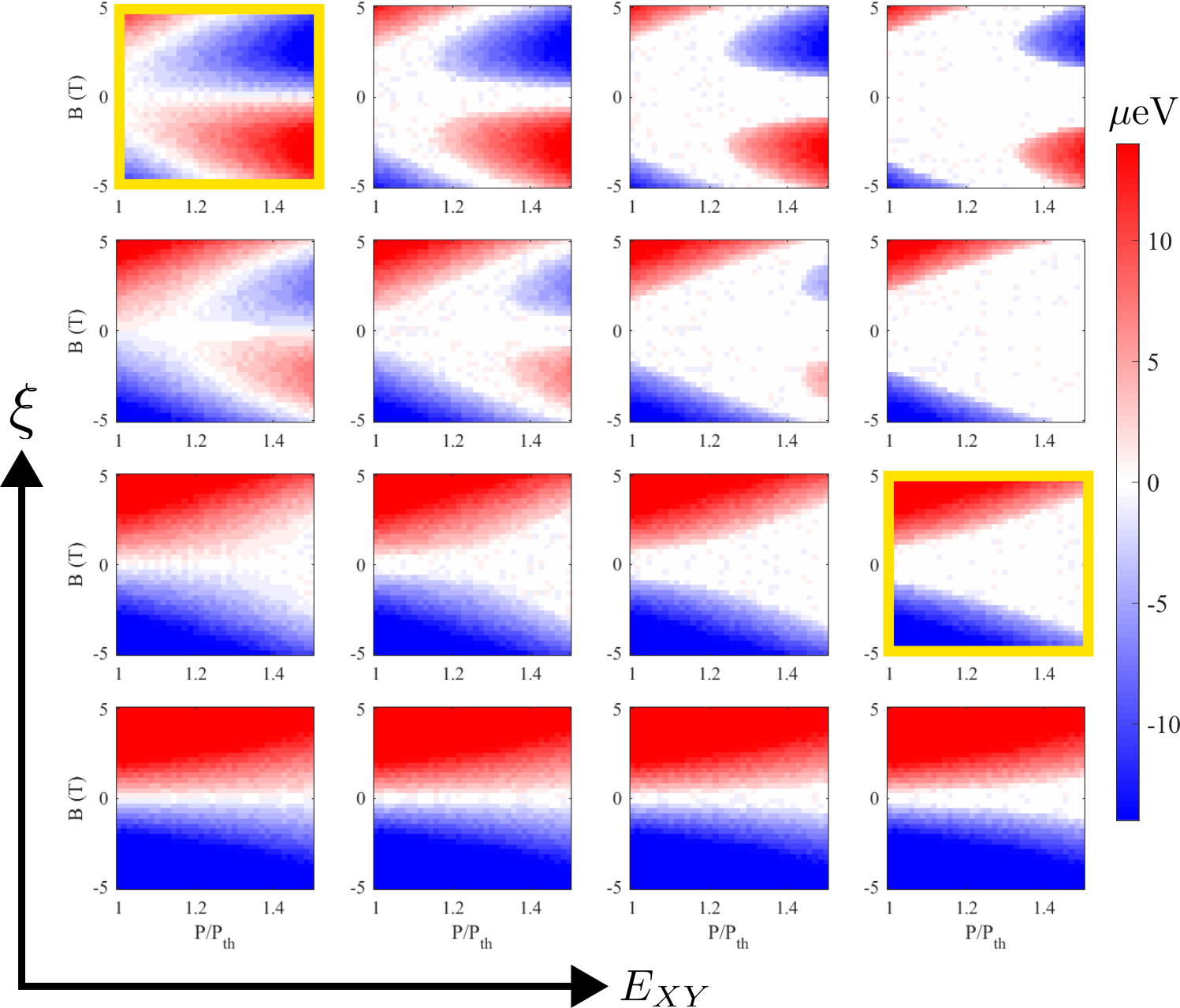}
\caption{\label{Fig10} Numerically obtained energy-difference between the polariton spin components from solving the 0DGPE in time. Panels from left to right: $E_{XY} = (0, 0.2, 0.4, 0.6)$ ns$^{-1}$. Panels from bottom to top: $\xi = (17,27,37,47)$. The two yellow outlined panels correspond approximately to Fig.~\ref{KSawicki_Fig4_suppression}(b) and~\ref{KSawicki_Fig5_inversion}(b).}
\end{figure*}

\section{\label{APPENDIX_C_parameters} Parameters of simulations}
Parameters to obtain the result presented in Fig.~\ref{KSawicki_Fig4_suppression}(b): $\xi=20$, $E_{\mathrm{ph}}- E_{\mathrm{X},0}=-1.9\,\mathrm{meV}$, $E_{\mathrm{XY}}=0.4\,\mu\mathrm{eV}$, $\Omega_{0} = 4\,\mathrm{meV}$, $u_\mathrm{X}=10$ $\mu\mathrm{eV}$, $R=1.2\times10^{-2}\,\mathrm{ps}^{-1}$, $\gamma^{-1} = 5.5\,\mathrm{ps}$, $\Gamma=2W=5\Gamma_{\mathrm{s}} = 5\times10^{-2}\,\mathrm{ps}^{-1}$, $a_{0}=10\,\mathrm{nm}$, and $\eta_1=2 \times 10^{-4}\,\mathrm{ps}^{-1}\,\mathrm{T}^{-1}$.

Parameters to obtain the result presented in Fig.~\ref{KSawicki_Fig5_inversion}(b) and Fig.~\ref{KSawicki_Fig7_analytical} are the same as in Fig.~\ref{KSawicki_Fig4_suppression}(b) except: $\xi=47$, and $E_{\mathrm{XY}}=0$.

\section{Below-threshold solution} \label{AppendixD}
Below the condensation threshold, we have $n_\pm >0$ and $\rho_\pm = 0$. The Zeeman splitting felt by any uncondensed polaritons is then modified by the reservoir as follows,
\begin{equation}
E_{\mathrm{ZS}} = E_{\mathrm{ZS}}^{(0)} + G_{+}\left(n_+ + \frac{\mathcal{P}_+}{W} \right) -G_{-}\left(n_- + \frac{\mathcal{P}_-}{W} \right).
\label{eq2_zeeman_splitting}
\end{equation}
The steady state solution of~\eqref{reservoir_rate_equation}, for linearly polarized pumping, results in the following expression for the reservoir contribution to the different blueshifts of the polariton spins below threshold,
\begin{align} \label{eq.nX}
\begin{split}
    & n_\sigma + \frac{\mathcal{P}_\sigma}{W}  = \frac{P}{W+2\Gamma_s} \bigg[ \frac{1}{2} + \frac{\Gamma_{s,-\sigma}}{W}  \\ 
    & +\frac{(\Gamma + \Gamma_{s,-\sigma})(\frac{W}{2}+\Gamma_{s,-\sigma}) + \Gamma_{s,-\sigma}(\frac{W}{2}+\Gamma_{s,\sigma})}{\Gamma(\Gamma+2\Gamma_s)} \bigg].
\end{split}
\end{align}
Here, we have used $\sigma = \pm$ for brevity.

\section{Energy splitting as a function of $E_{XY}$ and $\xi$} \label{AppendixE}
Figure~\ref{Fig10} shows the numerically obtained energy splitting of the condensate spins when varying $\xi$ (vertical) and $E_{XY}$ (horizontal). We see that the inversion is more pronounced when $\xi$ is increased. This is because the condensate is able to counter better against the real magnetic field and flipping the precession of the pseudospin. When $E_{XY}$ is increased a fixed point attractor becomes more pronounced corresponding to a larger white area (i.e., screened condensate). The panels with the yellow borders correspond to Figs.~\ref{KSawicki_Fig4_suppression} and~\ref{KSawicki_Fig5_inversion}. Each pixel in the numerical results is corresponds to a simulation starting from a random initial conditions which explains the slight noise in the obtained energies.

\section*{Acknowledgments}
This work was supported by the European Union Horizon 2020 program, through a Future and Emerging Technologies (FET) Open research and innovation action under Grant Agreement No. 964770 (TopoLight) and No. 899141 (PoLLoC). Y.W.’s studentship was financed by the Royal Society, Grant No. RGF$\backslash$EA$\backslash$180062. 
H.S. acknowledges the project No. 2022/45/P/ST3/00467 co-funded by the Polish National Science Centre and the European Union Framework Programme for Research and Innovation Horizon 2020 under the Marie Skłodowska-Curie grant agreement No. 945339.


\end{document}